\documentclass[aps,pra,twocolumn,amsmath,amssymb,nofootinbib,showpacs,superscriptaddress]{revtex4-1}
\usepackage[english]{babel}
\usepackage{latexsym}
\usepackage{graphics}
\usepackage{graphicx}
\usepackage{epsfig}
\usepackage{color}
\usepackage{bm}
\usepackage{amsmath}
\usepackage{amssymb}
\usepackage{amsthm}
\usepackage{dcolumn}
\usepackage{bm}
\usepackage{float}
\usepackage{hyperref}
\usepackage{color}
\usepackage{epstopdf}
\usepackage{cleveref}
\usepackage[svgnames]{xcolor}
\hypersetup{hidelinks,colorlinks=true,allcolors=DarkBlue}

\newcommand{\ket}[1]{|#1\rangle}
\newcommand{\bra}[1]{\langle#1|}

\begin{document}

\preprint{APS/123-QED}

\title{Teleportation in an indivisible quantum system}
\author{E.O. Kiktenko}
\affiliation{Theoretical Department, DEPHAN, Skolkovo, Moscow 143025, Russia}
\affiliation{Bauman Moscow State Technical University, Moscow 105005, Russia}
\author{A.K. Fedorov}
\affiliation{Theoretical Department, DEPHAN, Skolkovo, Moscow 143025, Russia}
\affiliation{Russian Quantum Center, Skolkovo, Moscow 143025, Russia}
\affiliation{LPTMS, CNRS, Univ. Paris-Sud, Universit\'e Paris-Saclay, Orsay 91405, France}
\author{V.I. Man'ko}
\affiliation{P.N. Lebedev Physical Institute, Russian Academy of Sciences, Moscow 119991, Russia}
\affiliation{Moscow Institute of Physics and Technology (State University), Moscow Region 141700, Russia}

\date{\today}

\begin{abstract}
Teleportation protocol is conventionally treated as a method for quantum state transfer between two spatially separated physical carriers.
Recent experimental progress in manipulation with high-dimensional quantum systems opens a new framework for implementation of teleportation protocols.
We show that the one-qubit teleportation can be considered as a state transfer between subspaces of the whole Hilbert space of an indivisible eight-dimensional system.
We explicitly show all corresponding operations and discuss an alternative way of implementation of similar tasks.

\begin{description}
\item[PACS numbers]
03.65.Wj, 03.65.-w, 03.67.-a
\end{description}
\end{abstract}

\maketitle

Intriguing phenomena of quantum teleportation \cite{Wootters} is an important ingredient for quantum technologies  
such as building of quantum networks \cite{Zoller,Gisin,Kimble} and large-scale quantum computers \cite{Chuang}.
Thanks to achievement of high level control for individual quantum systems \cite{Lukin},
quantum teleportation has been demonstrated in experiments with a variety of quantum agents, {\it e.g.},
photons \cite{Zeilinger}, 
atomic ensembles \cite{Blatt,Wineland,Polzik,Polzik2},
and superconducting circuits \cite{Wallraff,Wallraff2}.

In the framework of quantum information theory, 
teleportation is a noise-free identity channel from one Hilbert space of quantum states to the other \cite{Holevo}.
Conventionally, these Hilbert spaces are associated with two distinct locally separated physical objects.
Then teleportation is not related to transportation of physical systems, but to transfer of their quantum states \cite{Nielsen}.

However, this is not a dogma that Hilbert spaces in teleportation protocols should be associated to different systems.
Indeed, for realization of teleportation protocol one can use a decomposition of an indivisible quantum system on virtual subsystems \cite{MAManko}, 
that is allowed by representation of the Hilbert space of the indivisible system as a tensor product of low-dimensional Hilbert spaces.

Recently, indivisible quantum systems have attracted a great deal of interest both in theory and experiments.
Indivisible quantum systems have been significantly studied with
photons \cite{Padua}, 
ions \cite{Hensinger}, 
NMR \cite{Gedik}, 
and superconducting circuits \cite{DodonovManko,Dodonov,Katz,Katz2,Gustavsson,Katz3,Ustinov}.
It has been demonstrated theoretically that indivisible quantum systems can be used in quantum key distribution protocols \cite{QKD1,QKD2,QKD3,QKD4}, 
information processing \cite{Kessel,Kessel2,Cereceda,Luo,Gedik,Kiktenko,Kiktenko2},
and other algorithms \cite{QBC,Tavakoli}.
Information and entropic properties, 
in particular, schemes for verification of entropic inequalities \cite{Kiktenko, Glushkov}, 
of indivisible quantum systems have been investigated \cite{MAManko,Chernega,MAManko2,MAManko3}. 

In the present note, we demonstrate that the single-qubit quantum teleportation protocol can be employed in an indivisible quantum systems.
In this case, 
quantum teleportation can be considered as a state transfer between subspaces of the Hilbert space of the indivisible system in the same way as 
it commonly treated as state transfer in composite three-qubit system. 
We explicitly show all corresponding operations with the system.
The suggested scheme can be realized in eight-level quantum systems such as superconducting artificial atoms or spin-7/2 particles. 
We note that our approach has much in common with schemes for the quantum teleportation with identical particles \cite{Marzolino1,Marzolino2}.

\begin{figure}
\includegraphics[width=0.65\linewidth]{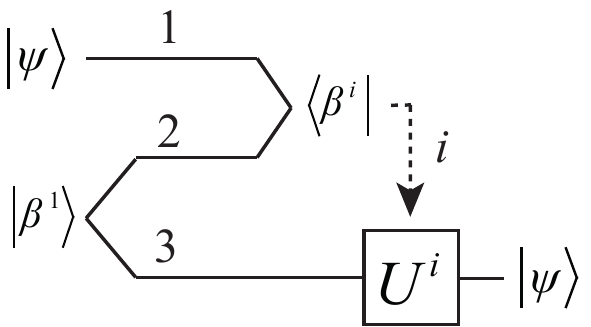}
\vskip -3mm
\caption
{
	Scheme of the one-qubit teleportation protocol via three distinct particles.
}
\label{fig:1}
\end{figure}

\begin{table*}[htp] 
\begin{tabular} {| c | } 
\hline 
$A^1{=}\begin{bmatrix}
	1 & 0 & 0 & 0 & 0 & 0 & 1 & 0 \\
	0 &	1 & 0 & 0 & 0 & 0 & 0 & 1 \\
	0 &	0 & 0 & 0 & 0 & 0 & 0 & 0 \\
	0 &	0 & 0 & 0 & 0 & 0 & 0 & 0 \\
	0 &	0 & 0 & 0 & 0 & 0 & 0 & 0 \\
	0 &	0 & 0 & 0 & 0 & 0 & 0 & 0 \\
	1 & 0 & 0 & 0 & 0 & 0 & 1 & 0 \\
	0 &	1 & 0 & 0 & 0 & 0 & 0 & 1 \\
\end{bmatrix}$
$A^2{=}\begin{bmatrix}
	0 &	0 & 0 & 0 & 0 & 0 & 0 & 0 \\
	0 &	0 & 0 & 0 & 0 & 0 & 0 & 0 \\
	0 &	0 & 1 & 0 & 1 & 0 & 0 & 0 \\
	0 &	0 & 0 & 1 & 0 & 1 & 0 & 0 \\
	0 &	0 & 1 & 0 & 1 & 0 & 0 & 0 \\
	0 &	0 & 0 & 1 & 0 & 1 & 0 & 0 \\
	0 &	0 & 0 & 0 & 0 & 0 & 0 & 0 \\
	0 &	0 & 0 & 0 & 0 & 0 & 0 & 0 \\
\end{bmatrix}$
$A^3{=}\begin{bmatrix}
	1 & 0 & 0 & 0 & 0 & 0 & -1 & 0 \\
	0 &	1 & 0 & 0 & 0 & 0 & 0 & -1 \\
	0 &	0 & 0 & 0 & 0 & 0 & 0 & 0 \\
	0 &	0 & 0 & 0 & 0 & 0 & 0 & 0 \\
	0 &	0 & 0 & 0 & 0 & 0 & 0 & 0 \\
	0 &	0 & 0 & 0 & 0 & 0 & 0 & 0 \\
	-1 & 0 & 0 & 0 & 0 & 0 & 1 & 0 \\
	0 &	-1 & 0 & 0 & 0 & 0 & 0 & 1 \\
\end{bmatrix}$
$A^4{=}\begin{bmatrix}
	0 &	0 & 0 & 0 & 0 & 0 & 0 & 0 \\
	0 &	0 & 0 & 0 & 0 & 0 & 0 & 0 \\
	0 &	0 & 1 & 0 & -1 & 0 & 0 & 0 \\
	0 &	0 & 0 & 1 & 0 & -1 & 0 & 0 \\
	0 &	0 & -1 & 0 & 1 & 0 & 0 & 0 \\
	0 &	0 & 0 & -1 & 0 & 1 & 0 & 0 \\
	0 &	0 & 0 & 0 & 0 & 0 & 0 & 0 \\
	0 &	0 & 0 & 0 & 0 & 0 & 0 & 0 \\
\end{bmatrix}$
\\
\hline
$B^1{=}\begin{bmatrix}
	1 & 0 & 0 & 0 & 0 & 0 & 0 & 0 \\
	0 & 1 & 0 & 0 & 0 & 0 & 0 & 0 \\
	0 & 0 & 1 & 0 & 0 & 0 & 0 & 0 \\
	0 & 0 & 0 & 1 & 0 & 0 & 0 & 0 \\
	0 & 0 & 0 & 0 & 1 & 0 & 0 & 0 \\
	0 & 0 & 0 & 0 & 0 & 1 & 0 & 0 \\
	0 & 0 & 0 & 0 & 0 & 0 & 1 & 0 \\
	0 & 0 & 0 & 0 & 0 & 0 & 0 & 1 
\end{bmatrix}$
$B^2{=}\begin{bmatrix}
	0 & 1 & 0 & 0 & 0 & 0 & 0 & 0 \\
	1 & 0 & 0 & 0 & 0 & 0 & 0 & 0 \\
	0 & 0 & 0 & 1 & 0 & 0 & 0 & 0 \\
	0 & 0 & 1 & 0 & 0 & 0 & 0 & 0 \\
	0 & 0 & 0 & 0 & 0 & 1 & 0 & 0 \\
	0 & 0 & 0 & 0 & 1 & 0 & 0 & 0 \\
	0 & 0 & 0 & 0 & 0 & 0 & 0 & 1 \\
	0 & 0 & 0 & 0 & 0 & 0 & 1 & 0
\end{bmatrix}$
$B^3{=}\begin{bmatrix}
	1 & 0 & 0 & 0 & 0 & 0 & 0 & 0 \\
	0 & -1 & 0 & 0 & 0 & 0 & 0 & 0 \\
	0 & 0 & 1 & 0 & 0 & 0 & 0 & 0 \\
	0 & 0 & 0 & -1 & 0 & 0 & 0 & 0 \\
	0 & 0 & 0 & 0 & 1 & 0 & 0 & 0 \\
	0 & 0 & 0 & 0 & 0 & -1 & 0 & 0 \\
	0 & 0 & 0 & 0 & 0 & 0 & 1 & 0 \\
	0 & 0 & 0 & 0 & 0 & 0 & 0 & -1 
\end{bmatrix}$
$B^4{=}\begin{bmatrix}
	0 & 1 & 0 & 0 & 0 & 0 & 0 & 0 \\
	-1 & 0 & 0 & 0 & 0 & 0 & 0 & 0 \\
	0 & 0 & 0 & 1 & 0 & 0 & 0 & 0 \\
	0 & 0 & -1 & 0 & 0 & 0 & 0 & 0 \\
	0 & 0 & 0 & 0 & 0 & 1 & 0 & 0 \\
	0 & 0 & 0 & 0 & -1 & 0 & 0 & 0 \\
	0 & 0 & 0 & 0 & 0 & 0 & 0 & 1 \\
	0 & 0 & 0 & 0 & 0 & 0 & -1 & 0
\end{bmatrix}$
\\
\hline
\end{tabular} 
\caption
{
	Explicit form of matrices $\{A_i\}$ and unitary operators $\{B_i\}$ in the teleportation protocol (\ref{eq:indtelep}) inside an indivisible eight-level system.
} 
\label{tab:ops} 
\end{table*}

First of all, we review briefly the standard teleportation protocol \cite{Wootters}.
We start from consideration of a set of maximally entangled Bell basis states
\begin{equation} \label{eq:Bell}
\begin{aligned}
	&\ket{\beta^1}=2^{-1/2}(\ket{0}\otimes\ket{0}+\ket{1}\otimes\ket{1}), \\
	&\ket{\beta^2}=2^{-1/2}(\ket{0}\otimes\ket{1}+\ket{1}\otimes\ket{0}), \\
	&\ket{\beta^3}=2^{-1/2}(\ket{0}\otimes\ket{0}-\ket{1}\otimes\ket{1}), \\
	&\ket{\beta^4}=2^{-1/2}(\ket{0}\otimes\ket{1}-\ket{1}\otimes\ket{0}),
\end{aligned}
\end{equation}
where the upper indices numerate vectors in the basis,
and one-qubit arbitrary state 
\begin{equation} \label{eq:arbstate}
	\ket{\psi}=\alpha\ket{0}+\beta\ket{1}, \quad |\alpha|^2+|\beta|^2=1.
\end{equation}

One-qubit teleportation protocol implies three particles: 
the first being initialized in arbitrary state (\ref{eq:arbstate}) and two others  in of one the Bell states (\ref{eq:Bell}) (see Fig.~\ref{fig:1}). 
Without loss of generality, one can consider the case of $\ket{\beta^1}$.
Particles 1 and 2 go to the party traditionally named Alice, while the particle 3 goes to party named Bob. 

The initial state of the three-qubit system can be written in the following form
\begin{equation} \label{eq:inst}
	\rho^{\mathrm{in}}_{123}=\ket{\psi}_1\bra{\psi}\otimes\ket{\beta^1}_{23}\bra{\beta^1},
\end{equation}
where subindices are used for a definition of the corresponding particles.
The goal of the teleportation protocol is to transfer the state $\ket{\psi}$ on the Bob's particle 3.
The protocol consists of several steps.
First, Alice performs a projective measurement of her particles in the Bell basis (\ref{eq:Bell}).
Then Alice sends the index of the result (two bits of classical information) to Bob.
On the final stage, Bob applies a unitary operator to his particle 3, 
which depends on the message obtained from Alice in the previous step.

The sequence of these operations yields the following state of particles:
\begin{equation}\label{eq:Telep}
\begin{split}
\rho^\mathrm{out}_{123}&=\sum_{i=1}^{4}\left(\ket{\beta^i}_{12}\bra{\beta^i}\otimes U^i_3\right)\rho^{\mathrm{in}}_{123} \left(\ket{\beta^i}_{12}\bra{\beta^i}\otimes {U^i_3}^\dagger\right)=\\
&=\frac{\mathbf{1} ^4_{12}}{4}\otimes\ket{\psi}_3\bra{\psi},
\end{split}
\end{equation}
where Bob's unitary operators are as follows:
\begin{equation}
\begin{split}
	U^1&=\mathbf{1}^2 \quad U^2=\sigma^x \\ 
	U^3&=\sigma^z \quad U^4=i\sigma^y.
\end{split}
\end{equation}
Here, $\mathbf{1}^n$ is $n$-dimensional identity operator, $\sigma^x$, $\sigma^y$ and $\sigma^z$ are standard Pauli operators.

Taking the partial trace over particles 1 and 2, we obtain particle 3 in the initial state of particle 1:
\begin{equation}
	\mathrm{Tr}_{12}\rho^{\mathrm{out}}_{123}=\ket{\psi}_3\bra{\psi},
\end{equation}
that is the essence of quantum teleportation protocol.

In the described above setup, there is a spacial separation between particles carrying the corresponding states of qubits.
However, if the spatial distance between qubits tends to zero one has an analog of a localized high dimensional system that can be simulated by a single qudit.

In what follows, we reformulate the teleportation protocol for an indivisible system of higher dimension. 
More precisely, we consider the teleportation protocol in a eight-dimensional Hilbert space, 
which can be always expressed as a tensor product of three two-dimensional subspaces:
\begin{equation}
	\mathcal{H}=\mathcal{H}_1\otimes\mathcal{H}_2\otimes\mathcal{H}_3,
\end{equation}
where $\mathcal{H}_1$, $\mathcal{H}_2$ and $\mathcal{H}_3$ do not correspond to any distinct physical carriers. 

For the encoding of three-qubit states in the Hilbert space of eight-dimensional system we consider the mapping
\begin{equation}
	\ket{n} \leftrightarrow \ket{a}_1\otimes \ket{b}_2\otimes\ket{c}_3,
\end{equation}
where $\{\ket{n}\}_{n=0}^{7}$ forms an orthonormal basis in $\mathcal{H}$ 
(that can be energy eigenstates or spin-7/2 projections on a particular axis) and $(abc)$ is binary expression of $n$ (\emph{e.g.} $n=3$ corresponds to $a=0$, $b=1$, $c=1$).

In line with Eq. (\ref{eq:inst}), 
the initial state of the system in $\mathcal{S}(\mathcal{H})$,
where $\mathcal{S}$ is a set of positive-semidefinite unit-trace operators in the corresponding Hilbert space, 
takes the form:
\begin{equation}
\varrho^{\mathrm{in}}=\frac{1}{2}
	\begin{bmatrix}
		|\alpha|^2 & 0 & 0 & |\alpha|^2 & \alpha\beta^* & 0 & 0 & \alpha\beta^* \\
		0 & 0 & 0 & 0 & 0 & 0 & 0 & 0 \\
		0 & 0 & 0 & 0 & 0 & 0 & 0 & 0 \\
		|\alpha|^2 & 0 & 0 & |\alpha|^2 & \alpha\beta^* & 0 & 0 & \alpha\beta^* \\
		\alpha^*\beta & 0 & 0 & \alpha^*\beta & |\beta|^2 & 0 & 0 & |\beta|^2 \\
		0 & 0 & 0 & 0 & 0 & 0 & 0 & 0 \\
		0 & 0 & 0 & 0 & 0 & 0 & 0 & 0 \\
		\alpha^*\beta & 0 & 0 & \alpha^*\beta & |\beta|^2 & 0 & 0 & |\beta|^2 \\
	\end{bmatrix},
\end{equation}
where $\varrho$ in the notation is used to emphasize that we consider a state of single object.

Then, transformation (\ref{eq:Telep}) has the following form
\begin{equation}\label{eq:indtelep}
	\varrho^{\mathrm{out}}=\frac{1}{4}\sum_{i=1}^4 B^i A^i \varrho^{\mathrm{in}} A^i {B^i}^\dagger,
\end{equation}
where $A^i/2=\ket{\beta^i}\bra{\beta^i}\otimes\mathbf{1}^2$ is a set of rank-2 projectors and $B^i=\mathbf{1}^4\otimes U^i$ is a set of eight-dimensional unitary operators.

The final state obtains block diagonal form
\begin{equation}\label{eq:finstateind}
	\varrho^{\mathrm{out}}=\frac{1}{4}{\rm diag}(\sigma,\sigma,\sigma,\sigma),
	\quad
	\sigma=
	{\begin{bmatrix}
	|\alpha|^2 & \alpha\beta^* \\
	\alpha^*\beta & |\beta|^2
	\end{bmatrix}}.
\end{equation}
We provide an explicit form of all matrices in Tab.~\ref{tab:ops}.

Partial trace of the final state (\ref{eq:finstateind}) over the Hilbert spaces $\mathcal{H}_1$ and $\mathcal{H}_2$ 
gives the state (\ref{eq:arbstate}) in the Hilbert space $\mathcal{H}_3$, 
that now is a subspace of single indivisible system in similar way as $\mathcal{H}_1$ and $\mathcal{H}_2$.
The open question is how the operation (\ref{eq:indtelep}) is related to applying the SWAP gate
\begin{equation}
\mathrm{SWAP}_{1\leftrightarrow 3}=\begin{bmatrix}
1 & 0 & 0 & 0 & 0 & 0 & 0 & 0 \\
0 & 0 & 0 & 0 & 1 & 0 & 0 & 0 \\
0 & 0 & 1 & 0 & 0 & 0 & 0 & 0 \\
0 & 0 & 0 & 0 & 0 & 0 & 1 & 0 \\
0 & 1 & 0 & 0 & 0 & 0 & 0 & 0 \\
0 & 0 & 0 & 0 & 0 & 1 & 0 & 0 \\
0 & 0 & 0 & 1 & 0 & 0 & 0 & 0 \\
0 & 0 & 0 & 0 & 0 & 0 & 0 & 1 \\
\end{bmatrix},
\end{equation}
which implements a similar task by exchanging states from $\mathcal{S}(\mathcal{H}_1)$ and $\mathcal{S}(\mathcal{H}_3)$.
However, the acting of SWAP gate and teleportation in an indivisible system are different.
On the one hand, the latter requires preparation of the Bell state in $\mathcal{H}_2\otimes\mathcal{H}_3$, 
but on the other hand it yields maximally mixed states in $\mathcal{H}_1\otimes\mathcal{H}_2$, that can be important for quantum information processing.

Consideration of this question in more details could provide interesting insights for relation between teleportation and computation in the framework of indivisible quantum systems, 
{\it e.g}, multilevel artificial atoms realized by multilevel superconducting circuits.

Finally we would like to note, that the presented concept of teleportation in an in indivisible quantum system is close to the teleportation with identical particles, 
where modes instead of particles are teleported~\cite{Marzolino1,Marzolino2}.
Due to indistinguishability there are no observables that act on a single particle leaving unchanged all the others, and in some sense we can think about them as a single object.
Such systems can be a promising resource for quantum computation in addition to multi-qubit systems, where an each qubit is realized by distinct physical object.

We are grateful to U. Marzolino for valuable comments. 
Authors thank anonymous referee for constructive comments.
The support by the Ministry for Education and Science of the Russian Federation within the framework of the Federal Program under Contract 14.579.21.0104 is gratefully acknowledged.

~
~
~
~
~


\begin{thebibliography}{}

\bibitem{Wootters}
C.H. Bennett, G. Brassard, C. Cr\'epeau, R. Jozsa, A. Peres, and W.K. Wootters,
{\href{http://dx.doi.org/10.1103/PhysRevLett.70.1895}{Phys. Rev. Lett. {\bf 70}, 1895 (1993)}};
for a review, see 
S. Pirandola, J. Eisert, C. Weedbrook, A. Furusawa, and S.L. Braunstein,
{\href{http://dx.doi.org/10.1038/nphoton.2015.154}{Nat. Photonics {\bf 9}, 641 (2015)}}.

\bibitem{Zoller}
H.-J. Briegel, W. D\"ur, J.I. Cirac, and P. Zoller,
{\href{http://dx.doi.org/10.1103/PhysRevLett.81.5932}{Phys. Rev. Lett. {\bf 81}, 5932 (1998)}}. 

\bibitem{Gisin}
H. de Riedmatten, I. Marcikic, W. Tittel, H. Zbinden, D. Collins, and N. Gisin,
{\href{http://dx.doi.org/10.1103/PhysRevLett.92.047904}{Phys. Rev. Lett. {\bf 92}, 047904 (2004)}}. 

\bibitem{Kimble}
H.J. Kimble,
{\href{http://dx.doi.org/10.1038/nature07127}{Nature (London) {\bf 453}, 1023 (2008)}}.

\bibitem{Chuang}
D. Gottesman and I.L. Chuang,
{\href{http://dx.doi.org/10.1038/46503}{Nature (London) {\bf 402}, 390 (1999)}}.

\bibitem{Preskill}
D. Gottesman, A. Kitaev, and J. Preskill,
{\href{http://dx.doi.org/10.1103/PhysRevA.64.012310}{Phys. Rev. A {\bf 64}, 012310 (2001)}}.

\bibitem{Lukin}
J. Thompson and M.D. Lukin,
{\href{http://dx.doi.org/10.1126/science.1256529}{Science {\bf 345}, 272 (2014)}}.

\bibitem{Zeilinger}
D. Bouwmeester, J.-W. Pan, K. Mattle, M. Eibl, H. Weinfurter, and A. Zeilinger, 
{\href{http://dx.doi.org/10.1038/37539}{Nature (London) {\bf 390}, 575 (1997)}}.

\bibitem{Blatt}
M. Riebe, H. H\"affner, C.F. Roos, W. H\"ansel, J. Benhelm, G.P.T. Lancaster, T.W. K\"orber, C. Becher, F. Schmidt-Kaler, D.F.V. James, and R. Blatt,
{\href{http://dx.doi.org/10.1038/nature02570}{Nature (London) {\bf 429}, 734 (2004)}}.

\bibitem{Wineland}
M.D. Barrett, J. Chiaverini, T. Schaetz, J. Britton, W.M. Itano, J.D. Jost, E. Knill, C. Langer, D. Leibfried, R. Ozeri, and  D.J. Wineland,
{\href{http://dx.doi.org/10.1038/nature02570}{Nature (London) {\bf 429}, 737 (2004)}}.

\bibitem{Polzik}
H. Krauter, D. Salart,	 C.A. Muschik, J.M. Petersen, H. Shen, T. Fernholz, and E.S. Polzik, 
{\href{http://dx.doi.org/10.1038/nphys2631}{Nature Phys. {\bf 9}, 400 (2013)}}.

\bibitem{Polzik2}
J.F. Sherson, H. Krauter, R.K. Olsson, B. Julsgaard, K. Hammerer, I. Cirac, and E.S. Polzik,
{\href{http://dx.doi.org/10.1038/nature05136}{Nature (London) {\bf 443}, 557 (2006)}}.

\bibitem{Wallraff}
M. Baur, A. Fedorov, L. Steffen, S. Filipp, M.P. da Silva, and A. Wallraff, 
{\href{http://dx.doi.org/10.1103/PhysRevLett.108.040502}{Phys. Rev. Lett. {\bf 108}, 040502 (2012)}}.

\bibitem{Wallraff2}
L. Steffen,	Y. Salathe, M. Oppliger, P. Kurpiers, M. Baur, C. Lang, C. Eichler, G. Puebla-Hellmann, A. Fedorov, and A. Wallraff,
{\href{http://dx.doi.org/10.1038/nature12422}{Nature (London) {\bf 500}, 319 (2013)}}.

\bibitem{Holevo}
A.S. Holevo, 
{\it Quantum systems, channels, information. A mathematical introduction} 
(De Gruyter, Berlin--Boston, 2012).

\bibitem{Nielsen}
M.A. Nielsen and I.L. Chuang,
{\it Quantum computation and quantum information} 
(Cambridge University Press, 2000).

\bibitem{MAManko}
M.A. Man'ko and V.I. Man'ko, 
{\href{http://dx.doi.org/10.1088/1742-6596/538/1/012016}{J. Phys.: Conf. Ser. {\bf 538}, 012016 (2014)}}.

\bibitem{Padua}
L. Neves, G. Lima, J.G. Aguirre G\'omez, C.H. Monken, C. Saavedra, and S. P\'adua,
{\href{http://dx.doi.org/10.1103/PhysRevLett.94.100501}{Phys. Rev. Lett. {\bf 94}, 100501 (2005)}}. 

\bibitem{Hensinger}
X. Zhang, M. Um, J. Zhang, S. An, Y. Wang, D. Deng, C. Shen, L.-M. Duan, and Kihwan Kim,
{\href{http://dx.doi.org/10.1103/PhysRevLett.110.070401}{Phys. Rev. Lett. {\bf 110}, 070401 (2013)}};
J. Randall, S. Weidt, E.D. Standing, K. Lake, S.C. Webster, D.F. Murgia, T. Navickas, K. Roth, and W.K. Hensinger, 
{\href{http://dx.doi.org/10.1103/PhysRevA.91.012322}{Phys. Rev. A {\bf 91}, 012322 (2015)}}. 

\bibitem{Gedik}
Z. Gedik, I.A. Silva, B. \c{C}akmak, G. Karpat, E.L.G. Vidoto, D.O. Soares-Pinto, E.R. deAzevedo, and F.F. Fanchini,
{\href{http://dx.doi.org/10.1038/srep07982}{Sci. Rep. {\bf 5}, 14671 (2015)}}.

\bibitem{Katz}
Y. Shalibo, Y. Rofe, I. Barth, L. Friedland, R. Bialczack, J.M. Martinis, and N. Katz,
{\href{http://dx.doi.org/10.1103/PhysRevLett.108.037701}{Phys. Rev. Lett. {\bf 108}, 037701 (2012)}}.

\bibitem{Katz2}
Y. Shalibo, R. Resh, O. Fogel, D. Shwa, R. Bialczak, J.M. Martinis, and N. Katz,
{\href{http://dx.doi.org/10.1103/PhysRevLett.110.100404}{Phys. Rev. Lett. {\bf 110}, 100404 (2013)}}.

\bibitem{Gustavsson}
M.J. Peterer, S.J. Bader, X. Jin, F. Yan, A. Kamal, T. Gudmundsen, P.J. Leek, T.P. Orlando, W.D. Oliver, and S. Gustavsson,
{\href{http://dx.doi.org/10.1103/PhysRevLett.114.010501}{Phys. Rev. Lett. {\bf 114}, 010501 (2015)}}.

\bibitem{Katz3}
E. Svetitsky, H. Suchowski, R. Resh, Y. Shalibo, J.M. Martinis, and N. Katz,
{\href{http://dx.doi.org/10.1038/ncomms6617}{Nat. Comm. {\bf 5}, 5617 (2015)}}.

\bibitem{Ustinov}
J. Braum\"uller, J. Cramer, S. Schl\"or, H. Rotzinger, L. Radtke, A. Lukashenko, P. Yang, M. Marthaler, L. Guo, A.V. Ustinov, and M. Weides,
{\href{http://dx.doi.org/10.1103/PhysRevB.91.054523}{Phys. Rev. B {\bf 91}, 054523 (2015)}}.

\bibitem{DodonovManko}
V.V. Dodonov, V.I. Man'ko, and O.V. Man'ko, 
{\href{http://dx.doi.org/10.1007/BF01120338}{J. Sov. Laser Res. {\bf 10}, 413 (1989)}};
{\href{http://dx.doi.org/10.1007/BF00866257}{Meas. Tech. {\bf 33}, 102 (1990)}}; 
{\href{http://dx.doi.org/10.1007/BF01121108}{J. Sov. Laser Res. {\bf 13}, 196 (1992)}};
Proc. Lebedev Phys. Inst. {\bf 200}, 155 (1991);
{\it ibid} {\bf 205}, 217 (1993).

\bibitem{Dodonov}
V.V. Dodonov, 
{\href{http://dx.doi.org/10.1002/0471231479.ch7}{Adv. Chem. Phys. {\bf 119}, 309 (2001)}};
{\href{http://dx.doi.org/10.1088/0031-8949/2010/T140/014020}{Phys. Scr. {\bf 82}, 038105 (2010)}}.

\bibitem{QKD1}
D. Bruss and C. Macchiavello, 
{\href{http://dx.doi.org/10.1103/PhysRevLett.88.127901}{Phys. Rev. Lett. {\bf 88}, 127901 (2002)}}.

\bibitem{QKD2}
N.J. Cerf, M. Bourennane, A. Karlsson, and N. Gisin, 
{\href{http://dx.doi.org/10.1103/PhysRevLett.88.127902}{Phys. Rev. Lett. {\bf 88}, 127902 (2002)}}.

\bibitem{QKD3}
T. Durt, N.J. Cerf, N. Gisin, and M. Zukowski, 
{\href{http://dx.doi.org/10.1103/PhysRevA.67.012311}{Phys. Rev. A {\bf 67}, 012311 (2003)}}.

\bibitem{QKD4}
S.P. Kulik, S.N. Molotkov, and I.V. Radchenko,
{\href{http://dx.doi.org/10.1134/S0021364012170080}{JETP Lett. {\bf 96}, 336 (2012)}}.

\bibitem{Kessel}
A.R. Kessel and V.L. Ermakov
{\href{http://dx.doi.org/10.1134/1.568130}{JETP Lett. {\bf 70}, 61 (1999)}};
{\href{http://dx.doi.org/10.1134/1.568340}{{\it ibid}. {\bf 71}, 307 (2000)}}.

\bibitem{Kessel2}
A.R. Kessel and N.M. Yakovleva,
{\href{http://dx.doi.org/10.1103/PhysRevA.66.062322}{Phys. Rev. A {\bf 66}, 062322 (2002)}}. 

\bibitem{Cereceda}
J.L. Cereceda,
{\href{http://arxiv.org/abs/quant-ph/0407253v4}{arXiv:quant-ph/0407253 (2004)}}.

\bibitem{Luo}
M.-X. Luo, X.-B. Chen, Y.-X. Yang, and X. Wang,
{\href{http://dx.doi.org/10.1038/srep04044}{Sci. Rep. {\bf 4}, 4044 (2014)}}.

\bibitem{Kiktenko}
E.O. Kiktenko, A.K. Fedorov, O.V. Man'ko, and V.I. Man'ko, 
{\href{http://dx.doi.org/10.1103/PhysRevA.91.042312}{Phys. Rev A \textbf{91}, 042312 (2015)}}.

\bibitem{Kiktenko2}
E.O. Kiktenko, A.K. Fedorov, A.A. Strakhov, and V.I. Man'ko,
{\href{http://dx.doi.org/10.1016/j.physleta.2015.03.023}{Phys. Lett. A {\bf 379}, 1409 (2015)}}. 

\bibitem{QBC}
R.W. Spekkens and T. Rudolph, 
{\href{http://dx.doi.org/10.1103/PhysRevA.65.012310}{Phys. Rev. A {\bf 65}, 012310 (2002)}}.

\bibitem{Tavakoli}
A. Tavakoli, A. Cabello, M. \.Zukowski, and M. Bourennane,
{\href{http://dx.doi.org/10.1038/srep07982}{Sci. Rep. {\bf 5}, 7982 (2015)}}.

\bibitem{Glushkov}
E. Glushkov, A. Glushkova, and V.I. Manko, 
{\href{http://dx.doi.org/10.1007/s10946-015-9522-z}{J. Russ. Laser Res. {\bf 36}, 448 (2015)}};
{\href{http://arxiv.org/abs/1509.04341}{arXiv:1509.04341}}.

\bibitem{Chernega}
V.N. Chernega, O.V. Manko, and V.I. Manko,
{\href{http://dx.doi.org/10.1007/s10946-013-9367-2}{J. Russ. Laser Res. {\bf 34}, 383 (2013)}};
{\href{http://dx.doi.org/10.1007/s10946-014-9447-y}{{\it ibid}. {\bf 35}, 457 (2014)}}.

\bibitem{MAManko2}
M.A. Man'ko and V.I. Man'ko, 
{\href{http://dx.doi.org/10.1088/0031-8949/2014/T160/014030}{Phys. Scr. {\bf T160}, 014030 (2014)}}.

\bibitem{MAManko3}
M.A. Man'ko and V.I. Man'ko, 
{\href{http://dx.doi.org/10.1142/S0219749915600060}{Int. J. Quantum Inf. {\bf 12}, 1560006 (2014)}}.

\bibitem{Marzolino1}
U. Marzolino and A. Buchleitner,
{\href{http://dx.doi.org/10.1103/PhysRevA.91.032316}{Phys. Rev. A  {\bf 91}, 032316 (2015)}};
{\href{http://arxiv.org/abs/1512.02692}{arXiv:1512.02692}}.

\bibitem{Marzolino2}
See also U. Marzolino and A. Buchleitner,
{\href{http://dx.doi.org/10.1088/0953-4075/44/9/091001}{J. Phys. B: At. Mol. Opt. Phys. {\bf 44}, 091001 (2011)}};
{\href{http://dx.doi.org/10.1142/S0219749911008210}{Int. J. Quant. Inf. {\bf 9}, 1745 (2011)}};
{\href{http://dx.doi.org/10.1103/PhysRevA.85.042329}{Phys. Rev. A {\bf 85}, 042329 (2012)}}.

\end{thebibliography}
\end{document}